\documentstyle{article}
\begin{document}
\Large
\begin{center}
{\Large \bf Big-Bang Nucleosynthesis:\\
Linking Inner Space and outer Space\footnote{\large Text of a poster on the theme
``Great Discoveries in Astronomy in the Last 100 Years'' produced for the
APS centennial meeting.} \\}
\vskip .25in
S. Burles, K. M. Nollett, and M. S. Turner\\
University of Chicago
\end{center}

A series of nuclear reactions took place when the Universe was seconds
old and made the lightest elements in the periodic table.  The
successful predictions of big-bang nucleosynthesis make it a
cornerstone of the hot big-bang cosmology.  It also leads to the most
accurate determination of the baryon density of the Universe, provides
the linchpin in the case for the existence of nonbaryonic dark matter,
and permits the study of fundamental physics in regimes beyond the
reach of terrestrial laboratories.

\section*{A BRIEF HISTORY}

Two of the most pressing problems of the first half of
this century were the energy source of stars
and the origin of the chemical elements.  In the 1930s the
first puzzle was solved when Hans Bethe and others worked
out the nuclear reactions that power stars like the sun.
Nuclear physicists then turned to the stars to solve the second puzzle.
However, by the end of the 1930s, they were ready to
abandon them.  In 1938 von Weizs\"acker articulated the dominant
view:  ``$\cdots$ no element heavier than $^4$He can be built up to
any appreciable extent.  Therefore we must assume that the heavy
elements were built up before the stars $\cdots$''

In 1942 George Gamow began talking about the big-bang origin of the
elements.  In the 1948 paper that marks the beginning of big-bang
nucleosynthesis (BBN), Gamow, his student Ralph Alpher, and Hans Bethe
proposed that the periodic table was built up by neutron capture
minutes after the big bang.  Critical physics corrections made by
Chushiro Hayashi, Enrico Fermi and Anthony Turkevich led to the
seminal 1953 paper of Alpher, Robert Herman and James Follin that
described correctly the big-bang synthesis of large amounts of $^4$He
and little else.  As Fermi and Turkevich had pointed out, Coulomb
barriers and the lack of stable nuclei with mass 5 and 8 preclude
significant nucleosynthesis beyond $^4$He.  BBN required a hot
beginning, and in 1949 Alpher and Herman predicted a 5 K temperature
for the relic radiation now known as the Cosmic Microwave Background
(CMB).

In 1957 Fred Hoyle, E. Margaret and Geoffrey Burbidge, and William
Fowler and independently A.G.W. Cameron showed that essentially all of
the elements beyond $^4$He can be made in stars, but under very
different conditions than von Weizs\"acker and others considered.
Interestingly enough, Hoyle was impelled to work on stellar
nucleosynthesis because of his attachment to the steady-state
cosmology which lacked an explosive beginning.  Years later, Hermann
Bondi, another father of the steady state, referred to the work of
Hoyle and collaborations as the most important achievement of the
steady-state theory.

A year before the discovery of the CMB, Hoyle and Roger Tayler made
the observational case for a large primeval abundance of $^4$He
(around 25\% by mass) and suggested a big-bang explanation.  After the
discovery of the CMB by Penzias and Wilson in 1965, the BBN
calculations were refined by P.J.E. Peebles and Robert Wagoner, Fowler
and Hoyle.  Explaining the large primeval abundance of $^4$He was a
striking triumph of the hot big-bang theory.  In 1973 Hubert Reeves,
Jean Audouze, Fowler and David N. Schramm focused attention on
deuterium, whose big-bang production depends sensitively on the
density of ordinary matter (baryons).  The reasoning in this paper
together with the detection of deuterium in the interstellar medium
(ISM) led to an upper limit to the baryon density of no more than 10\%
of the critical density.

By the early 1980s, primordial abundances of all four light elements
cooked in the big bang, D, $^3$He, $^4$He and $^7$Li, had been
determined, and the concordance of the predicted abundances with the
measured abundances was used both as a test of the big-bang framework
and a means of constraining the baryon density.  The hot big-bang
model had become the standard cosmology and BBN was one of its
cornerstones.  Further, BBN began to play an important role in probing
fundamental physics.  In their influential 1977 paper, Gary Steigman,
David Schramm and James Gunn used big-bang $^4$He production to
constrain the number of light neutrino species.  Not only was their
limit an important one, but it also helped to open the field of
cosmology and particle physics.

In 1998 the first accurate measurement of the primordial deuterium abundance
by David Tytler and Scott Burles marked the beginning of a new,
precision era of BBN.  This measurement fixes the baryon density to
a precision of around 7\%, and in turn, leads to accurate
predictions of the primeval abundances of the other light elements.
This development promises to extend and
sharpen the power of BBN to probe cosmology, fundamental physics and
astrophysics.

\newpage

\section*{HOW IT WORKS}

Big-bang nucleosynthesis is very different from the stellar
nucleosynthesis that produces the heavier elements.  It is a
nonequilibrium process that took place over the course of a few
minutes in an expanding, radiation-dominated plasma with high entropy
($10^9$ photons per baryon) and lots of free neutrons.  In contrast,
much of stellar nucleosynthesis occurs in equilibrium over billions of
years at relatively low entropy (less than one photon per baryon) and
no free neutrons.  The densities in stars are around $10^{2}\,{\rm
g\,cm^{-3}}$, while in the big bang they were closer $10^{-5}\,{\rm
g\,cm^{-3}}$.  The theoretical description of BBN requires only a few
basic assumptions -- general relativity, the standard big-bang
cosmology, and the standard model of particle physics -- along with a
dozen nuclear cross sections which are well measured at the relevant
energies.

At times much less than one second after the beginning, the Universe
was a hot ($\gg 10^{10}$\,K), rapidly expanding plasma, with most of
its energy in radiation and relativistic particles.  In particular,
there were roughly equal numbers of electrons, positrons, neutrinos
and antineutrinos (of each species), and photons.  Nucleons were
outnumbered by more than a billion to one.  There were essentially no
composite nuclei, and weak processes like $\nu + n \leftrightarrow p+
e^-$ maintained the ratio of neutrons to protons at its thermal
equilibrium value of around unity.

At about one second, the temperature had dropped to around $10^{10}$
K.  The weak processes became ineffective, and the neutron/proton
ratio leveled off at about $1/6$.  Growing amounts of D, $^3$He,
$^3$H, and $^4$He were present in amounts dictated by nuclear
statistical equilibrium.  The processes maintaining this equilibrium
slowed relative to the temperature evolution (because of decreasing
temperatures and densities).  After five minutes, most neutrons were
in $^4$He nuclei, and most protons remained free (see Figure).  Much
smaller amounts of D, $^3$He, and $^7$Li were synthesized, but the low
density, growing Coulomb barriers, and stability gaps at masses five
and eight worked against the formation of larger nuclei.  The
elemental composition of the Universe subsequently remained unchanged
until the formation of the first stars several billion years later.
The yields of primordial nucleosynthesis, with $2\sigma$ theoretical
errors, are shown as a function of the baryon density in the central
Figure.

\newpage
\section*{OBSERVATIONS CONFRONT PREDICTIONS}

The big-bang predictions for the light-element abundances
depend only upon the mean baryon density.
The primeval abundances of the four light elements are
not measured easily or simultaneously.  Here is a brief summary.

\begin{enumerate}

\item{\bf Helium-4}:  Since the big bang its abundance has
grown because stars make $^4$He.  The primordial abundance
is inferred from measurements of the $^4$He/H ratio in regions of hot,
ionized gas (HII regions) in other galaxies.
The Figure shows a compilation of these measurements
as a function of the Oxygen abundance, an indicator of stellar
processing.   Izotov and Thuan infer $Y_P = 0.244 \pm 0.002$.

\item{\bf Deuterium}: It is the most fragile of the light elements --
all astrophysical processes destroy D -- and so its
abundance has been declining since the big bang.  In 1973 Rogerson and York
measured the deuterium abundance in the local ISM; this measurement
provided a lower limit to the big-bang production and an upper limit
to the baryon density.  In 1998, Tytler and Burles
determined the primeval deuterium abundance
by measuring the D/H ratio in several high ($z>3$)
redshift hydrogen clouds, (D/H)$_P=(3.4\pm 0.3)\times 10^{-5}$.
These hydrogen clouds are ``seen'' by their distinctive Ly-$\alpha$
absorption features in the spectra of QSOs, with the deuterium
feature isotopically shifted (to the blue)
by $0.33(1+z_{\rm cloud})\,$\AA\
(see Figure).  The primeval deuterium abundance pins down the
baryon density, $(3.6\pm 0.2)\times 10^{-31}\,{\rm g\,cm^{-3}}$.

\item{\bf Lithium}: Some stars destroy lithium and others produce it.
The primeval value is inferred from the $^7$Li abundance in the
atmospheres of the oldest (pop II) stars in the halo of our galaxy,
($^7$Li/H)$_P= (1.7\pm 0.15)\times 10^{-10}$.  While the less massive
halo stars have depleted some of their lithium, the ``lithium
plateau'' for stars with higher surface temperatures suggests that
these stars have not (see Figure).  However, stellar models indicate
that there could have been up to a factor of two depletion on the
lithium plateau.

\item{\bf Helium-3}: Stars burn primeval deuterium to $^3$He; beyond
that little is certain.  It has been argued that the net destruction
or production beyond this is small.  If so, then the sum of D + $^3$He
remains relatively constant (measurements support this idea).  Under
this assumption, the primeval deuterium abundance together with the
measured abundance of D + $^3$He in the ISM imply that ($^3$He/H)$_P =
(0.3\pm 1)\times 10^{-5}$.

\end{enumerate}

Since the 1980s cosmologists have spoken of a concordance interval for
the baryon density where the predicted and measured abundances for all
four light elements are consistent (within their uncertainties).
Because the abundances span nine orders of magnitude, this is no mean
feat, and it establishes the validity of the standard cosmology when
the Universe was seconds old and a billion times smaller.  The
accurate determination of the primeval deuterium abundance changed the
strategy.  It pegged the baryon density, and led to accurate
predictions for the other light elements.  When the $^4$He abundance
is known better, a comparison with the predicted abundance,
$Y_P=0.246\pm 0.001$, will be an important consistency test.  When the
issue of stellar depletion is settled and the theoretical errors are
reduced, lithium will offer a similar test.  On the other hand, $^3$He
will serve best to probe of galactic and stellar evolution.  The
central Figure summarizes the present situation, showing concordance
intervals for each element (based upon $2\sigma$ uncertainties) and
the baryon density predicted by the deuterium measurement (vertical
band).  A remarkable cross check of BBN will be possible when
precision measurements of CMB anisotropy made by the MAP and Planck
satellites determine the baryon density to similar accuracy, based
upon the completely independent physics of gravity-driven acoustic
oscillations (see Figure).

\newpage

\section*{PROBING COSMOLOGY AND PARTICLE PHYSICS}

\subsection*{Baryonic and Nonbaryonic Dark Matter}

For more than a decade, the ``BBN concordance interval'' has stood as
the most accurate determination of the baryon density.  The
measurement of the primordial D abundance ushered in a new level of
precision: expressed as a fraction of the critical density, the baryon
density is $\Omega_B = 0.043 \pm 0.003$ (for $H_0=65\,{\rm km\,
sec^{-1}\,Mpc^{-1}}$ -- it varies as $H_0^{-2}$).  This has several
important implications: first, since stars contribute a mass density
that is about ten times less, it implies that most of the baryons must
be dark (most likely in the form of diffuse, hot gas).  Second,
measurements of the total matter density indicate that it is eight
times larger, $\Omega_M = 0.4\pm 0.1$.  BBN leads us to the remarkable
conclusion that most of the matter is something other than baryons.
The leading candidate is elementary particles (such as axions or
neutralinos) left over from the earliest, fiery moments.  While a
self-consistent picture of structure formation also argues for
nonbaryonic dark matter, BBN is truly the linchpin in the case.

\subsection*{Using Helium to Count Neutrinos}

The conditions at the time of big-bang nucleosynthesis are very
different than those available in terrestrial laboratories, and so it
is not surprising that BBN has been used as a heavenly laboratory to
probe physics in regimes that go beyond the reach of earthly
laboratories.  BBN has constrained the properties of neutrinos,
nucleons and nuclei, axions and other hypothetical particles, as well
as tested general relativity and the predictions of theories that
attempt to unify the forces and particles of Nature.  The most
striking example of the power of the heavenly laboratory is the
well-known limit to the number of neutrino species.

The physics works like this.  At the time of BBN, the energy density
of the Universe, which controls its expansion rate, was dominated by a
thermal bath of relativistic particles including neutrinos and
antineutrinos.  More neutrino species means a higher energy density
and faster expansion.  This leads to more neutrons, and hence more
$^4$He production (see Figure).  Thus, the $^4$He abundance can be
used to constrain the number of neutrino species and thereby the
number of families of quarks and leptons since there is one neutrino
for each family.  In 1977 when Steigman, Schramm and Gunn
obtained a limit of no more than 7 neutrino species, the direct
laboratory limit was around 5000 -- a truly impressive improvement.
By the time the $e^+e^-$ colliders at SLAC and CERN showed directly
that the number of neutrino species was 3, the BBN limit stood at no
more than 4.  Today, the BBN limit stands at no more than 3.2 (at
$2\sigma$) and is used to constrain the possible existence of other
new, light weakly interacting particles.

\section*{Selected Bibliography}

\begin{description}

\item{\bf Beginnings:}
G. Gamow, Phys. Rev. 22, 153 (1946);
R. Alpher, H. Bethe, and G. Gamow,
Phys. Rev. 73, 803 (1948)

\item{\bf Ready to Go:}
R. Alpher, J. Follin and R. Herman, Phys. Rev. 92, 1357 (1953)

\item{\bf The Helium Clue:}  F. Hoyle and R.J. Tayler,
Nature 203, 1108 (1964)

\item{\bf The Modern Era:}  R.V. Wagoner and D.N. Schramm,
Ann. Rev. Nuc. Sci. 27, 37 (1977)

\item{\bf The Concordance Era:}
A.M. Boesgaard and G. Steigman,
Ann. Rev. Astron. Astrophys. 23, 319 (1985)

\item{\bf The Precision Era:}  D.N. Schramm and M.S. Turner,
Rev. Mod. Phys. 70, 303 (1998); S. Burles, K. M. Nollett, J. W. Truran, and
M. S. Turner, Phys. Rev. Lett. (submitted), astro-ph/9901157

\end{description}

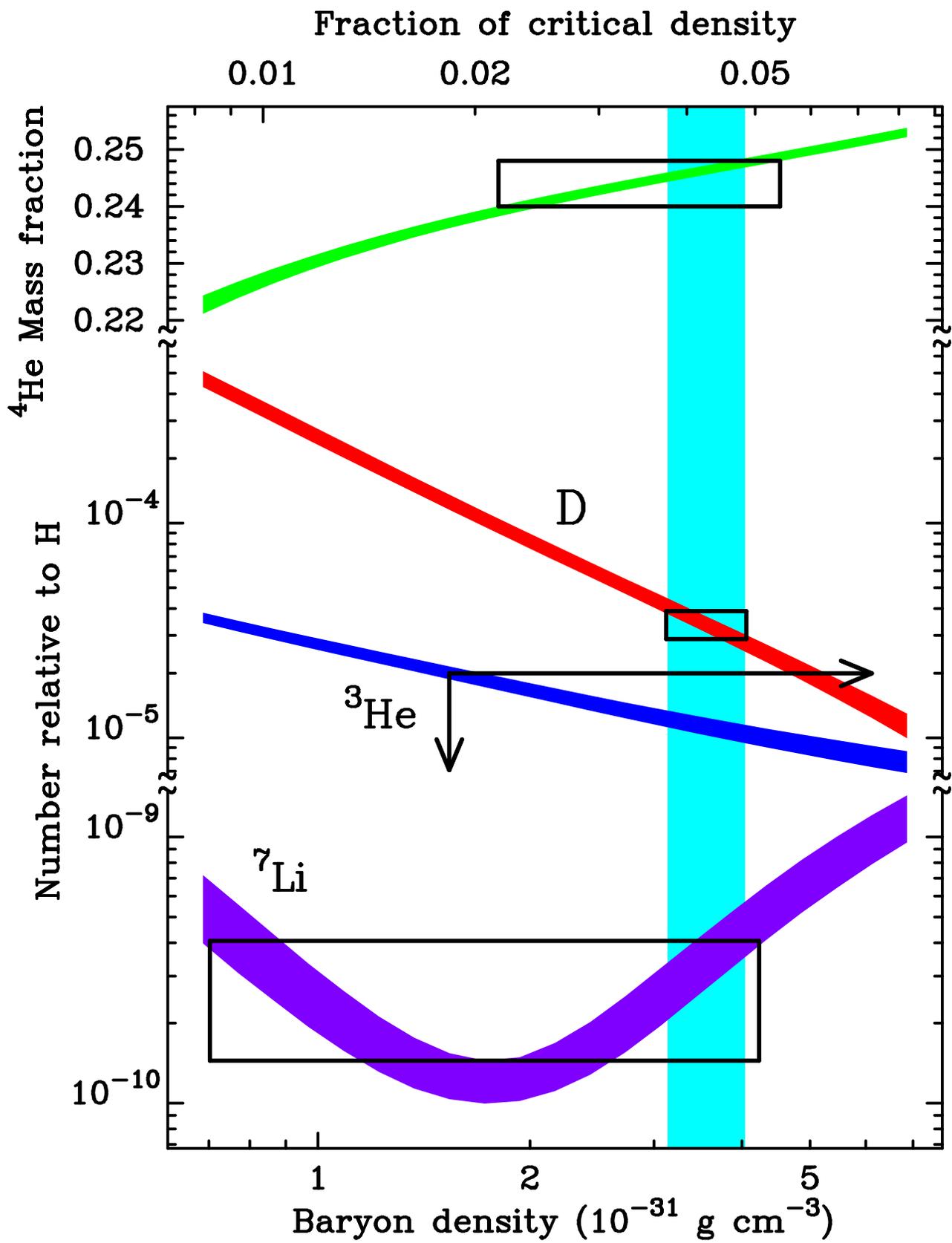
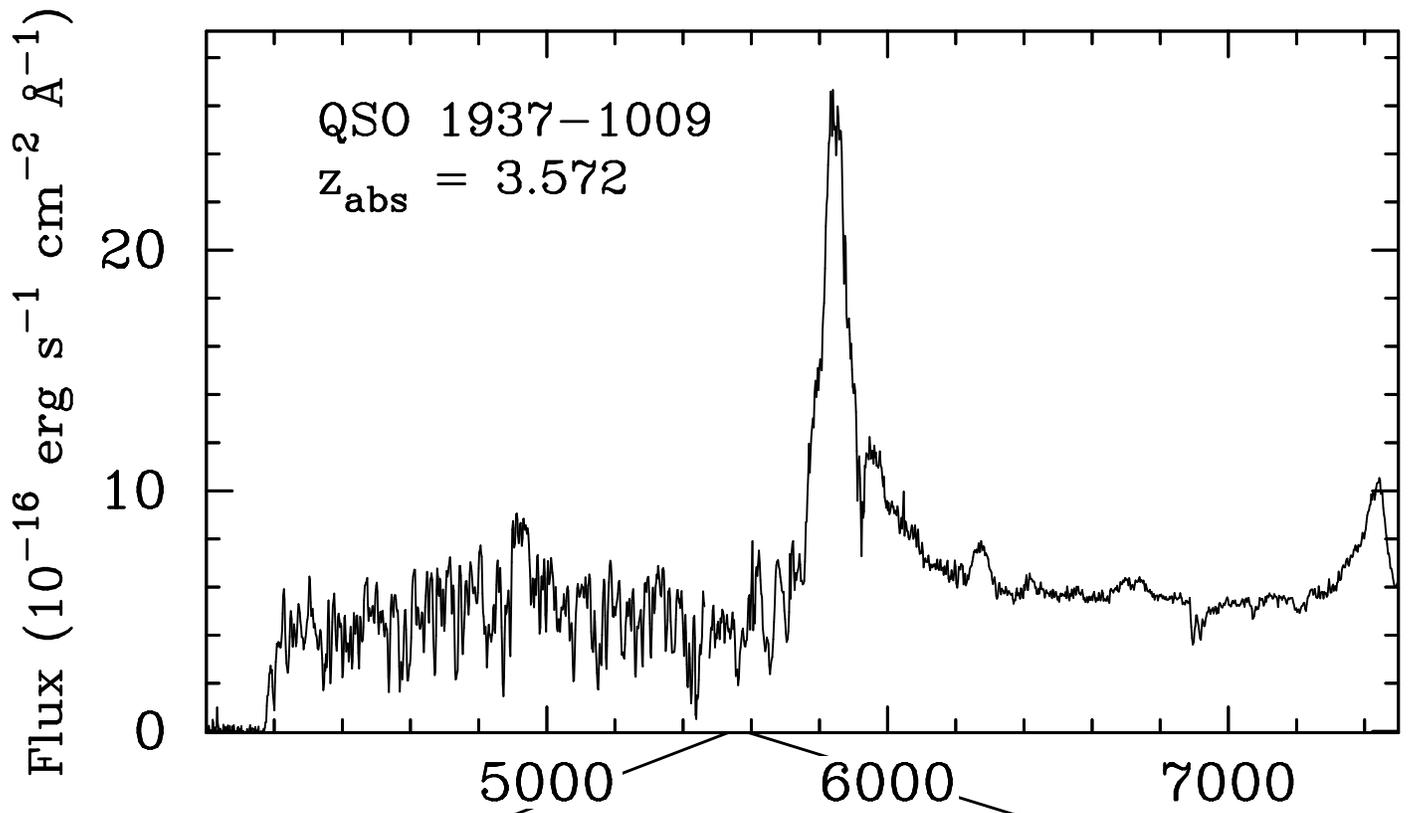
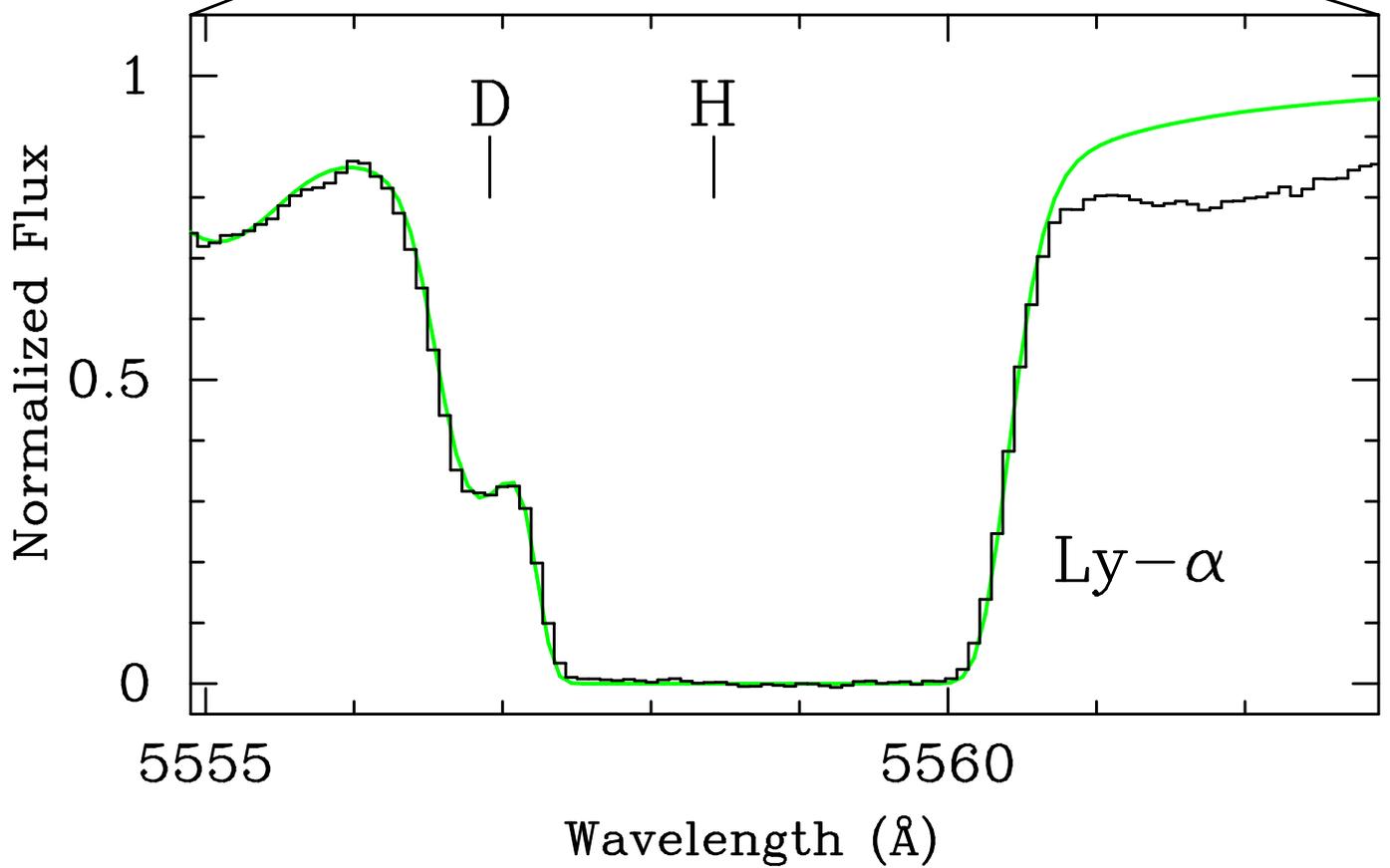
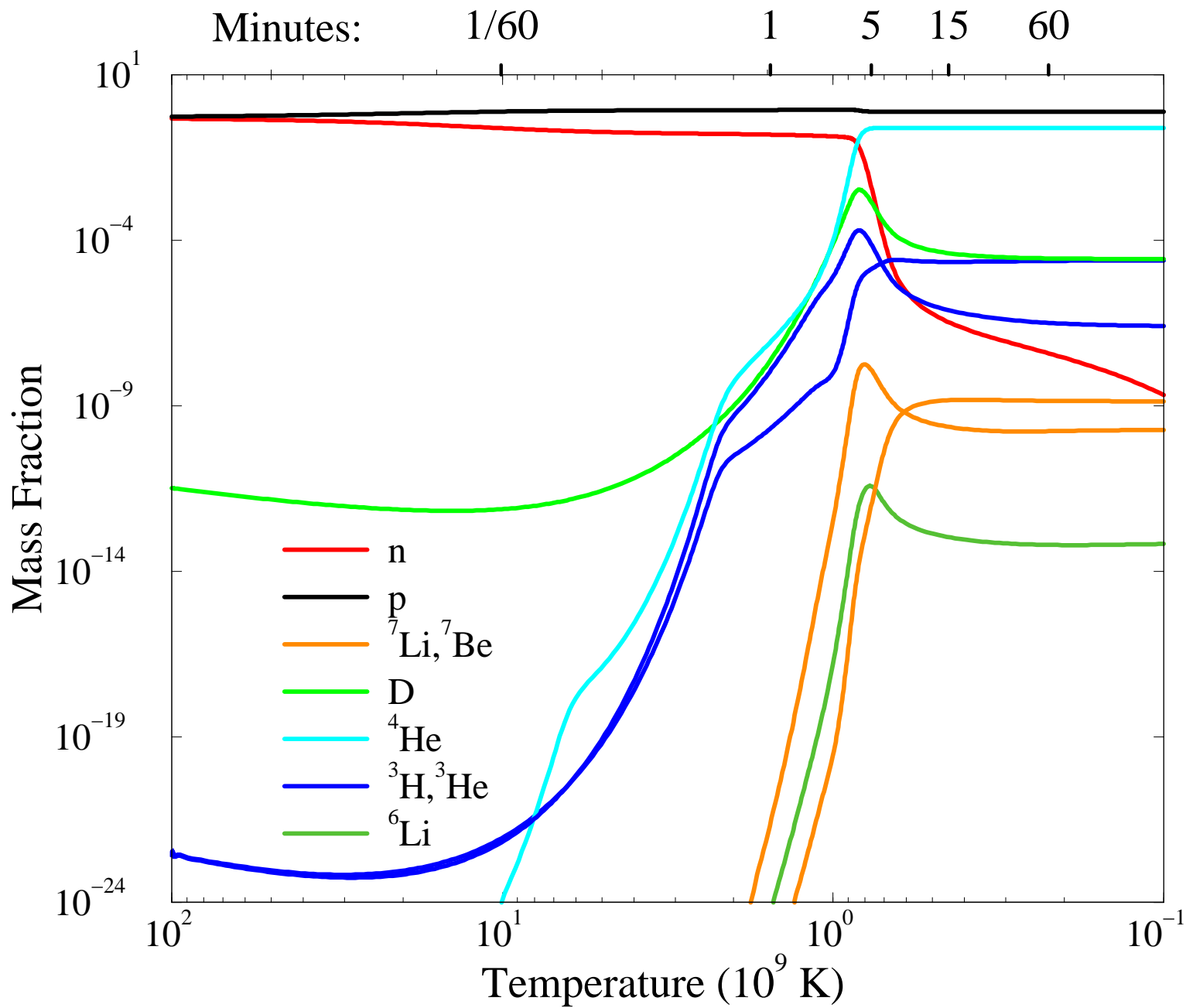
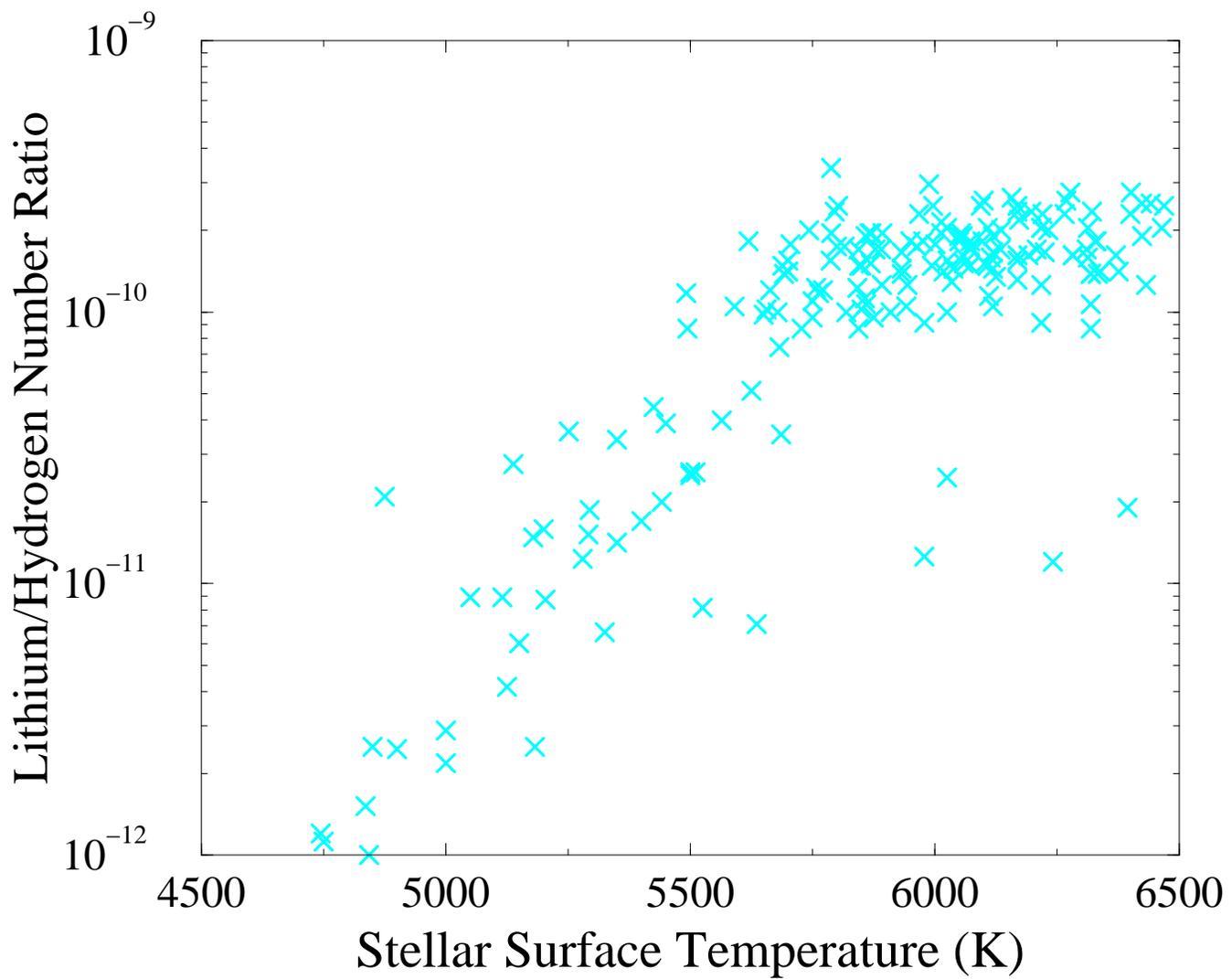
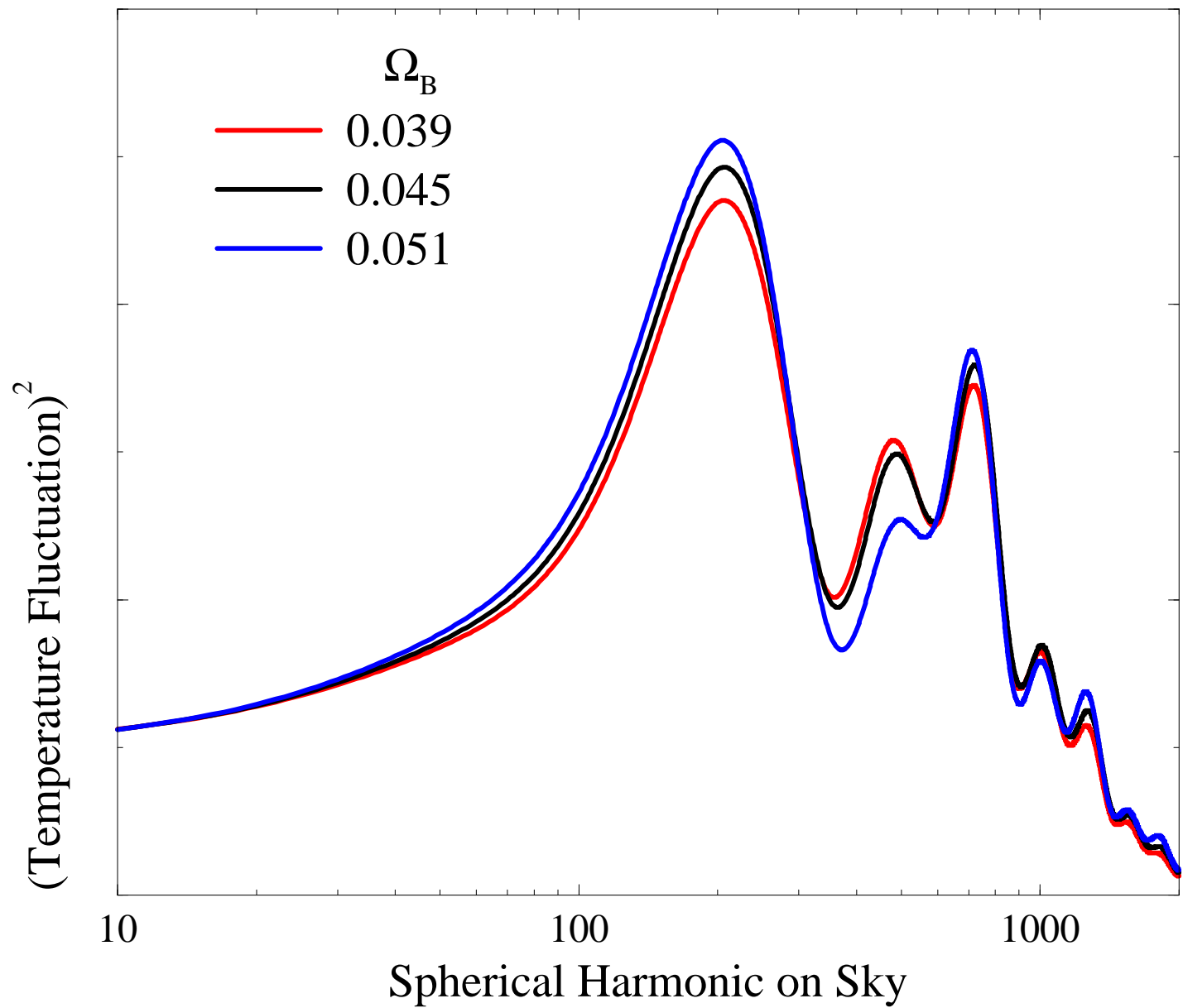
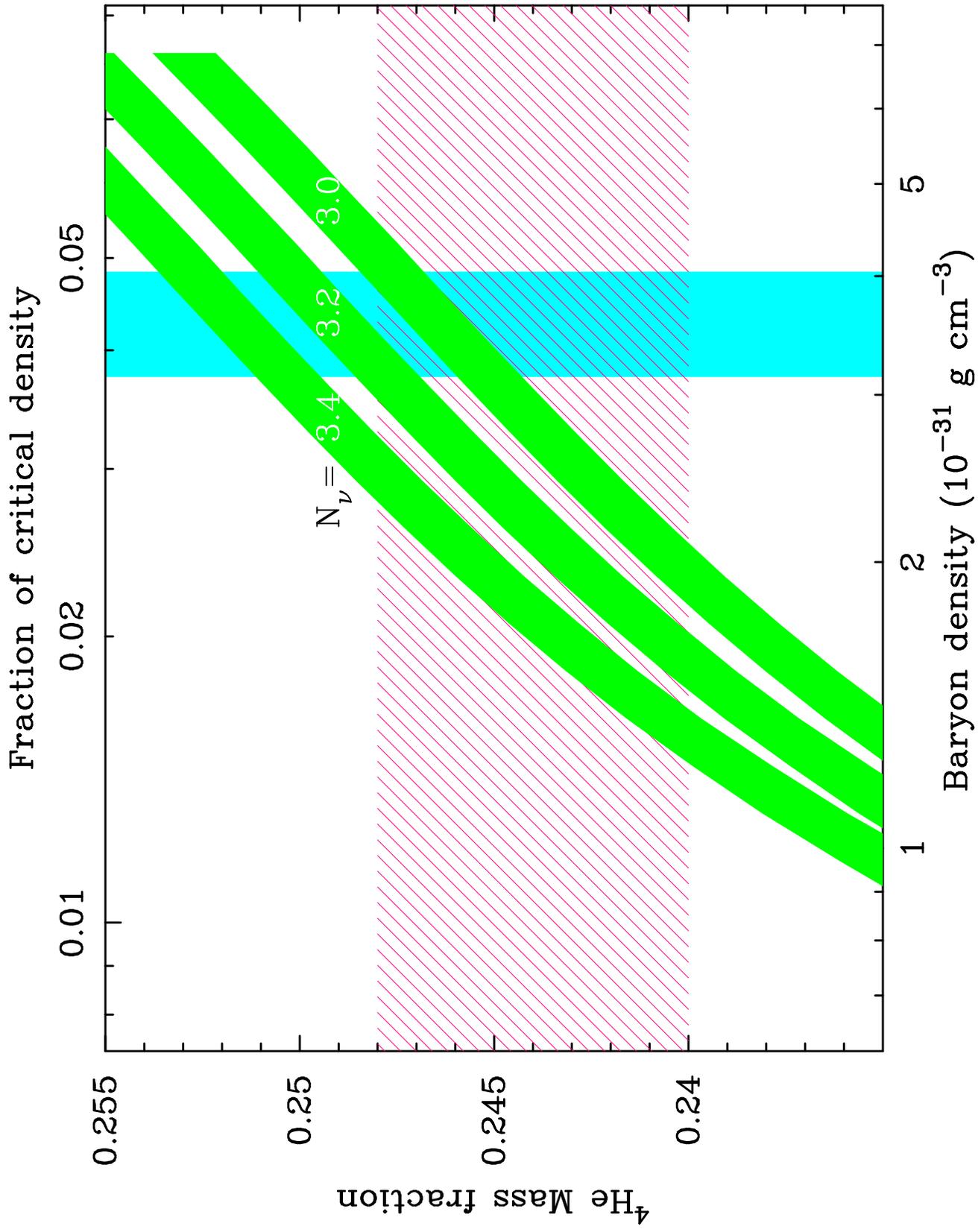
\end{document}